\documentstyle[b98proc,epsfig]{article}

\def\Journal#1#2#3#4{{#1} {\bf #2}, #3 (#4)}

\def\NPA{{\em Nucl. Phys.} A}

\def\PLB{{\em Phys. Lett.} B}

\def\PRL{\em Phys. Rev. Lett.}

\def\PRD{{\em Phys. Rev.} D}
\def\PRC{{\em Phys. Rev.} C}

\begin{document}

\title{
MUON CAPTURE AND THE PSEUDOSCALAR FORM FACTOR 
OF THE NUCLEON}

\author{V. BERNARD}
\address{ULP, Physique Th{\' e}orique, F-67084 Strasbourg}
\author{T. R. HEMMERT\footnote{Talk given at Baryons 98, Bonn, Sept 22-26, 1998, 
{\tt [FZJ-IKP(Th)-1998-30]} \\
email: th.hemmert@fz-juelich.de}, U.-G. MEI{\ss}NER}
\address{FZ J\"ulich, IKP (Theorie), J\"ulich, Germany}


\maketitle

\abstracts{We summarize recent work on muon capture and the pseudoscalar
form factor of the nucleon.}

\section{Introduction}
Ordinary ($\mu^- p\rightarrow \nu_\mu n$) and Radiative 
($\mu^- p\rightarrow \nu_\mu \gamma n$) Muon Capture (OMC, RMC) on a proton are
venerable subjects in nuclear physics ({\it e.g.} ref.\cite{Primakoff}). After 
having served
for decades as a testing ground for the symmetries and structure of the weak 
interaction, today these reactions can also be regarded as unique tests of the axial
structure of the nucleon as mandated by the explicitly and spontaneously broken 
chiral symmetry of QCD at low energies. In particular, they can give us access to the 
elusive
pseudoscalar form factor $G_P(q^2)$ of the nucleon which has received new 
attention~\cite{bkmgp,BFHM} many years after the pioneering analyses in the 
1960s~\cite{ADW}. Due to experimental constraints most of the muon capture work in
the single nucleon sector so far has focused on OMC. Therein one is only sensitive to one
particular kinematic point in the pseudoscalar form factor 
$\frac{m_\mu}{2 M_N}G_P(q^2=-0.88 m_\mu^2)\equiv g_p$, which is commonly referred
to as the pseudoscalar coupling constant. It took until 1995 that the first 
measurement of RMC on Hydrogen was reported~\cite{Triumf}, but quite surprisingly 
the extracted number for $g_P$ disagreed by as much as 50\%
from the very precise theoretical calculations ({\it e.g.} see \cite{ADW,bkmgp}).
In the following years RMC on the proton was reanalyzed in the framework of Heavy Baryon
Chiral Perturbation Theory (HBChPT) by several groups~\cite{chiralRMC}, but
as of September 1998 (i.e. BARYONS 98)
no new hadronic structure effect could be identified that would have invalidated
Fearing's calculation~\cite{Fearing} used in the analysis of the data. In parallel,
the chiral structure of $G_P(q^2)$ was reanalyzed~\cite{BFHM} and explicitly shown
to be unaffected by contributions from the first nucleon resonance $\Delta$(1232).
For now the discrepancy remains unexplained~\cite{Cheon}, 
with new theoretical~\cite{paper} 
and experimental~\cite{proposal} investigations under way.

In the remainder of this brief contribution we want to show that OMC and RMC indeed are 
low energy hadronic processes where a good convergence behavior of HBChPT can be 
expected~\cite{paper}. We then address a remaining open question in the existing 
RMC calculations. Furthermore, we emphasize the existence of very precise 
predictions~\cite{ADW,bkmgp,BFHM} for the momentum dependence of $G_P(q^2)$---which go 
beyond the usual focus on $g_p$ in the literature---and present the poor
state of ``world data'' for this ``black sheep'' among the nucleon form factors. 
Finally, we point out that pion electroproduction is a promising window to improve 
our knowledge of $G_P(q^2)$.

\section{Ordinary Muon Capture}

In the Fermi approximation of a static $W_\mu^-$ field,
the invariant matrix element of ordinary muon capture can be written as
\begin{eqnarray}
{\cal M}_{\mu^-p\rightarrow \nu_\mu n}&=&{\cal M}^{\rm OMC} =
                \langle \nu_\mu|W_\mu^+|\mu\rangle
                \,i\,\frac{g^{\mu\nu}}{M_W^2}\left[\langle n|V_\nu^-|p\rangle -
                \langle n|A_\nu^-|p\rangle\right]~.
\end{eqnarray}
The leptonic matrix element $\langle \nu_\mu|W_\mu^+|\mu\rangle$ is uniquely fixed
by the electroweak vertices of the Standard Model, utilized here as
the source of well-understood external fields that probe the hadronic structure of
a nucleon of mass $M_N$ and isovector magnetic moment $\mu_v$ at low energies.
Now one calculates the charge-changing hadronic vector 
$\langle n|V_\nu^-|p\rangle$ and axial-vector $\langle n|A_\nu^-|p\rangle$ 
currents in HBChPT to the order desired.
To ${\cal O}(p^2)$ one finds~\cite{paper}
\begin{eqnarray}
\langle n|V_\alpha^-|p\rangle^{(2)}&=&-i\frac{g_2 V_{ud}}{\sqrt{8}}
                             \bar{n}(r^\prime)\{
                             v_\alpha +\frac{(r+r^\prime)_\alpha}{2 M_N}
                             +\frac{\mu_v}{M_N}[S_\alpha ,S\cdot
                             (r^\prime-r)] \} p(r)\, ,  \nonumber \\
\langle n|A_\alpha^-|p\rangle^{(2)}&=&-i\frac{g_2 V_{ud}}{\sqrt{8}}
                             \bar{n}(r^\prime)\left\{
                             2\,g_AS_\alpha-\frac{g_A}{M_N}S\cdot(r+r^\prime)v_\alpha
                             \right. \label{eq3} \\
                          & &\phantom{-i\,\frac{g_2 V_{ud}}{\sqrt{8}}\,
                             \bar{n}(r^\prime)}\left.
                             -\frac{2g_AS\cdot(r^\prime-r)}{
                             (r^\prime-r)^2-m_{\pi}^2}
                             (r^\prime-r)_{\mu}\right\} p(r)+{\cal O}(1/M_N^2)
                             ~,\nonumber
\end{eqnarray}
where $v_\alpha,\,[S_\alpha]$ corresponds to the velocity- [spin-]vector of 
HBChPT
and $g_A,\,[g_2]$ denotes the axial vector [weak] coupling constant~\cite{paper} of the 
nucleon with CKM matrix element $V_{ud}$. Eq.(\ref{eq3}) contains the 
coupling of the axial 
source to the nucleon via an intermediate pion of mass $m_\pi$ as required by chiral
symmetry, leading to a ${\cal O}(p)$ effect in the transition current.

Assuming that the initial muon-proton system constitutes the ground-state
of a bound system described by a 1s Bohr-wavefunction $\Phi(x)_{1s}$ of a 
muonic atom one finds the 
{\em spin-averaged} capture rate ${\cal O}(p^2)$~\cite{paper}
\begin{eqnarray}
\Gamma_{\rm OMC}&=&\frac{\alpha^3 G_F^2 V_{ud}^2 m_\mu^5}{2\pi^2
               (m_\pi^2+m_\mu^2)^2} \left\{
               (2g_A^2+1)m_\mu^4+(4g_A^2+2
               )m_\mu^2m_\pi^2+(3g_A^2+1)m_\pi^4\right.\nonumber \\
            & &+\frac{2m_\mu}{(m_\pi^2+m_\mu^2)M_N}
               \left[(g_A\mu_v-5g_A^2-2
               )(m_\mu^6+3m_\mu^4m_\pi^2)\right.\nonumber \\
            & &\left.\left.+(3g_A\mu_v-16g_A^2-6)m_\mu^2m_\pi^4 
               +(g_A\mu_v-7g_A^2-2)m_\pi^6\right] \right\}
               +{\cal O}(1/M_N^2)\nonumber \\
            &=&\left( \,247\,-\,59\,\right)\times s^{-1}\,+{\cal O}(1/M_N^2) \, ,
               \label{gam}
\end{eqnarray}
with $G_F={g_2^2\sqrt{2}}/{(8 M_W^2)}$. Note that
the ${\cal O}(p^2)$ contribution amounts to a correction of less than
25\% of the leading term.
The expectation that OMC has a well behaved chiral expansion is 
also supported by the observation that in the case of no explicit chiral 
symmetry breaking (i.e. $m_\pi = 0$) the spin-averaged capture rate 
is only changed by 
10\%: $\Gamma_{\rm OMC}^{\chi}=(214-46)\times s^{-1}\,+{\cal O}(1/M_N^2)$.
The physical reason for the nice stability of perturbative calculations for OMC 
is of course
the fact that contributions of order $n$ are suppressed~\cite{paper} by 
$\left(m_i/\Lambda_\chi\right)^{n-1}$, with $i=\pi,\mu$ and 
$\Lambda_\chi\sim M_N\sim 1$GeV. Analogous suppression effects are at work for 
RMC~\cite{paper}. We therefore note that at ${\cal O}(1/M_N^2)$---when 
the calculation becomes sensitive to the internal structure of the 
nucleon beyond just the isovector magnetic moment $\mu_v$ and the leading pion pole of 
Eq.(\ref{eq3})---the new structure effects are strongly suppressed and therefore present
a formidable challenge for the required precision of muon capture 
experiments. 

\section{Some comments on RMC}

Several HBChPT calculations of RMC have been performed since BARYONS 95, the most 
elaborate one by Ando and Min~\cite{chiralRMC}. They found that the 
${\cal O}(p^3)$ effects are small, in accordance with the analysis presented in the 
previous section. No large structure effect that would invalidate the 
Born-term analysis of Fearing~\cite{Fearing} could be identified. However, in our
opinion there is one point left to be examined in detail regarding the
contribution of $\Delta$(1232) in muon capture. 
In HBChPT these effects are incorporated via
${\cal O}(p^3)$ counterterms, leading only to a small effect---consistent with
previous phenomenological analyses~\cite{BF}. While this result is reassuring
it is also surprising from the viewpoint of effective field theories. 
Introducing $\Delta$(1232) into the theory leads
to a new scale $\Delta=M_\Delta-M_N\sim 300$MeV, which would suggest that the resulting
effects $\left(m_i/\Delta\right),\,i=\pi,\mu$ could be of the 
order of 30\%! We have started
to investigate~\cite{paper} this problem to identify the origin of this strong 
suppression of $\Delta$(1232). First numerical results confirm the smallness of the 
contributions, but the {\em analytical} structure and the physics behind this suppression
is hard to pin down. It's origin lies in the fact that due to the atomic structure
of the $\mu p$ system both OMC and RMC are very sensitive to spin structure of the 
initial state which seems to act as filter mechanism~\cite{paper}.

\section{The elusive Form Factor}

The electroweak structure of a nucleon is typically encoded via
6 form factors ({\it e.g.} ref.\cite{BFHM}). Muon captures provides us with 
the opportunity to study the axial-weak structure of a nucleon. In the absence of 
second class currents the corresponding relativistic matrix element of
the hadronic axial current reads
\begin{eqnarray}
\langle n|A_\alpha^-|p\rangle=\bar{n}(p_2)\left[G_A (q^2)
\gamma_\alpha\gamma_5+\frac{G_P(q^2)}{2M_N}q_\alpha \gamma_5
\right]p(p_1).
\end{eqnarray}
Here, $G_A (q^2)$ and $G_P (q^2)$ are the axial and the induced
pseudoscalar form factor, respectively. While $G_A (q^2)$ can be
extracted from (anti)neutrino--proton scattering or charged pion
electroproduction data, $G_P(q^2)$ is harder to pin down and in fact 
{\em constitutes the least known nucleon form factor}. In Fig.1 we present the 
``world data'' for $G_P(q^2)$.
\begin{figure}[ht]
\vspace{5.5cm}
\includegraphics{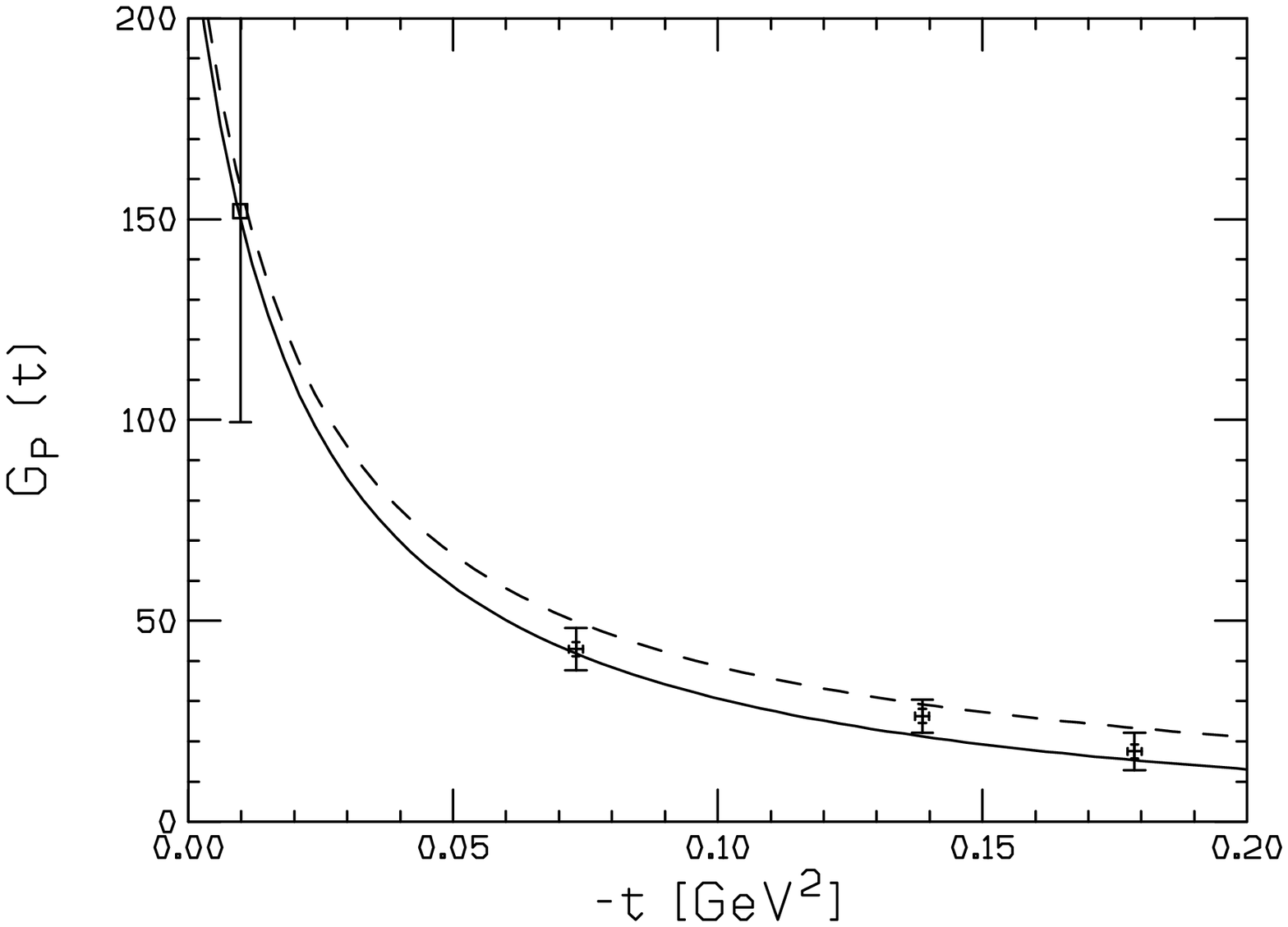}
\hspace{4.5cm}
\includegraphics{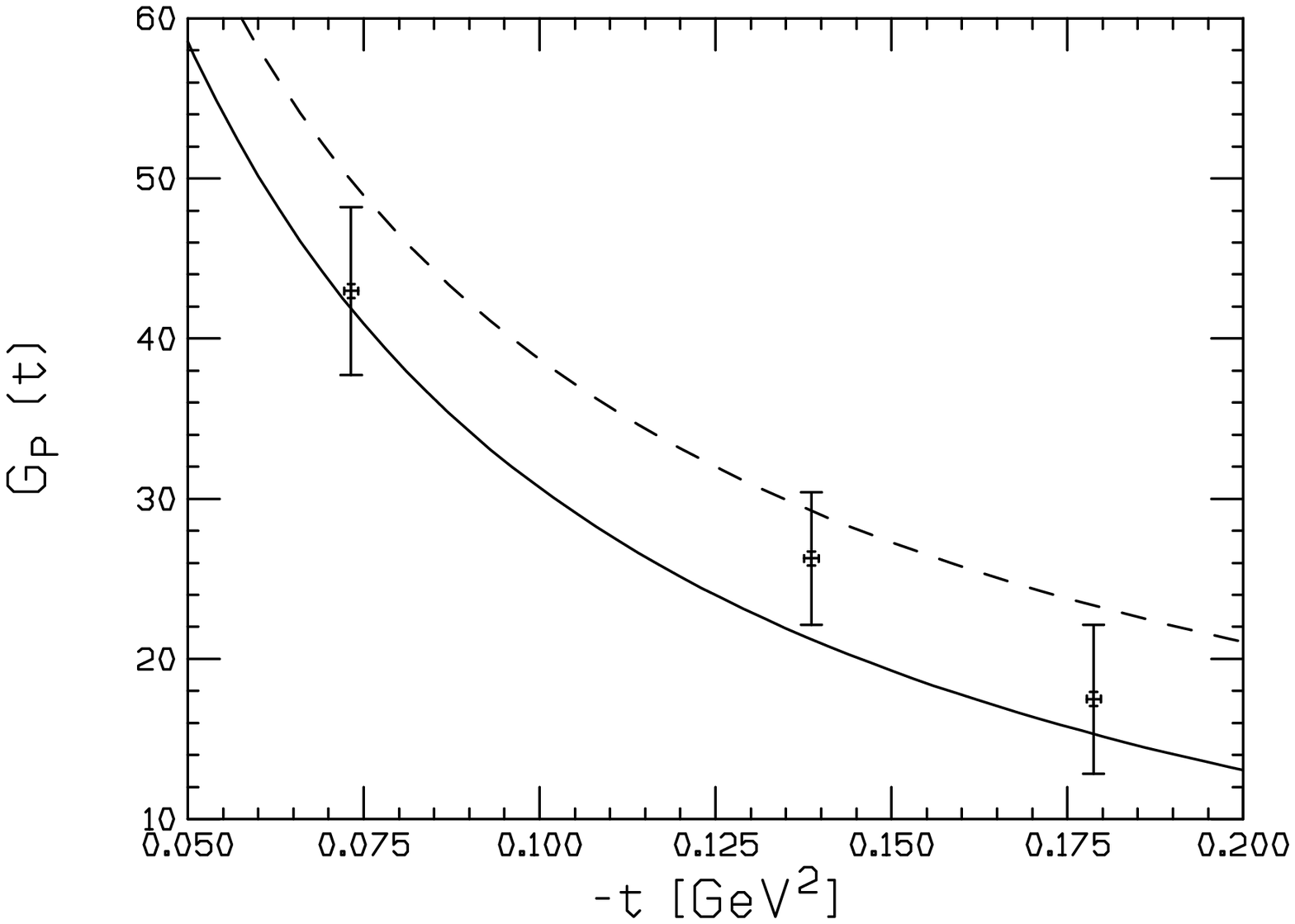}
\vspace{-2.2cm}
\caption{``World data'' for $G_P(t)$. 
Dashed curve: $\pi$-pole prediction. Solid curve:
Full chiral prediction; Right panel: Zoom of $\pi$-electroproduction region.}
\end{figure}
In OMC one is sensitive to the point
\begin{eqnarray}
g_P\equiv \frac{m_\mu}{2 M_N} \; G_P(q^2=-0.88 m_{\mu}^2) 
&=&\frac{2 m_\mu F_\pi g_{\pi NN}}{m_{\pi}^2+0.88 m_{\mu}^2}-\frac{1}{3}
      g_A m_\mu M_N r_{A}^2 \nonumber \\
   &=& 8.23\dots 8.46 \label{eq:gp} \; ,
\end{eqnarray}
with $F_\pi$ denoting the pion-decay constant and $r_A$ the axial radius of the nucleon 
extracted from $G_A(q^2)$.
It is this prediction for $g_p$ with which the RMC result
from TRIUMF~\cite{Triumf} disagrees, whereas the OMC~\cite{OMC} 
measurements are consistent with
it, within errors (see Fig.1). Note that the (theoretical) error of 
Eq.(\ref{eq:gp}) is much smaller and comes from the uncertainty in the strong 
coupling constant $g_{\pi NN}$. Eq.(\ref{eq:gp}) is obtained via several quite 
different theoretical analyses~\cite{ADW,bkmgp,BFHM} and nowadays is considered to 
rest on firm ground.
 
There are 2 curves shown in Fig.1 to 
display the difference between the usual pion-pole parameterization for $G_P(q^2)$
and analyses that take into account the full chiral structure of the form 
factor\cite{ADW,bkmgp,BFHM}, yielding
\begin{eqnarray}
G_{P}^{\chi}(q^2)&=&\frac{4 M_N g_{\pi NN} F_\pi}{m_{\pi}^2-q^2}-\frac{2}{3}
                   g_A M_{N}^2 r_{A}^2 \; . \label{eq:gpff}
\end{eqnarray}
In the kinematical region of RMC, which mainly lies to the
``left'' of the OMC point in Fig.1 the structure effect proportional to $r_A$ is expected
to play only a small role. Certainly, the present experimental uncertainties both in 
OMC~\cite{OMC} 
and in RMC~\cite{Triumf} are too large to distinguish between the 2 curves,
but new efforts are under way~\cite{proposal}.
Finally we want to emphasize that there exists another window on 
$G_P(q^2)$---pion electroproduction. So far there has only been one 
experiment~\cite{saclay} that
took up the challenge, with the results shown in Fig.1. {\em In this kinematical regime
the structure proportional to $r_A$ produces the biggest effect and a new dedicated 
experiment should be able to identify it}---thereby enhancing 
our knowledge of this poorly known form factor considerably!



\end{document}